\documentclass[aps,prb,twocolumn]{revtex4}
\usepackage{amsfonts}
\usepackage{amsmath}
\usepackage{graphicx}
\usepackage{dsfont}
\usepackage{diagbox}

\begin{document}

\title{Matrix product states for topological phases with parafermions}
\author{Wen-Tao Xu$^{1}$ and Guang-Ming Zhang$^{1,2}$}
\affiliation{$^{1}$State Key Laboratory of Low-Dimensional Quantum Physics and Department
of Physics, Tsinghua University, Beijing 100084, China. \\
$^{2}$Collaborative Innovation Center of Quantum Matter, Beijing 100084,
China.}
\date{\today}

\begin{abstract}
In the Fock representation, we propose a framework to construct the
generalized matrix product states (MPS) for topological phases with $\mathbb{%
Z}_{p}$ parafermions. Unlike the $\mathbb{Z}_{2}$ Majorana fermions, the $%
\mathbb{Z}_{p}$ parafermions form intrinsically interacting systems. Here we
explicitly construct two topologically distinct classes of irreducible $%
\mathbb{Z}_{3}$ parafermionic MPS wave functions, characterized by one or
two parafermionic zero modes at each end of an open chain. Their
corresponding parent Hamiltonians are found as the fixed point models of the
single $\mathbb{Z}_{3}$ parafermion chain and two-coupled parafermion chains
with $\mathbb{Z}_{3}\times \mathbb{Z}_{3}$ symmetry. Our results thus pave
the road to investigate all possible topological phases with $\mathbb{Z}_{p}$
parafermions within the matrix product representation in one dimension.
\end{abstract}

\maketitle

\section{Introduction}

Topological phases of matter have become one of the most important subjects
in condensed matter physics, because their low-energy excitations have
potential use for fault-tolerant quantum computation\cite{RMP}. Among them,
the simplest class is the symmetry protected topological (SPT) phases with
robust gapless edge excitations\cite%
{Chen-Gu-Wen-2011,Schuch,chen-gu-liu-wen,Chen-Gu-Wen-2011(2)}. Without
breaking the protecting symmetry or closing the energy gap, these SPT phases
can not be continuously connected to the trivial phase. In one dimension,
matrix product states (MPS) for bosonic SPT phases capture not only the
model dependent microscopic properties of quantum spin chain systems, but
also the universal properties associated to the family of Hamiltonians in
the same quantum phase\cite%
{Haldane2008,Pollman2010,Pollman2012,RaoWanZhang2014,Fu2014,RaoZhangYang2016}%
. In the valence-bond-solid picture\cite{AKLT}, the important feature of
these SPT phases is revealed as the edge particles with fractionalized
degrees of freedom, resulting in the degeneracy of the entanglement spectrum.

The fractionalized Majorana/parafermion zero modes also exist in fermionic
SPT phases and exhibit non-abelian statistics\cite%
{Kitaev2001,FidkowshiKitaev,Fendley2012}, however, it is not straight
forward to extend the matrix product representation to the class of
fermionic/parafermionic systems. Recently, it is noticed that the fermionic
MPS can be constructed by using the language of super vector spaces\cite%
{Bultinck2017,Kapustin2016}, where the basis states have a well-defined
parity of the fermion number. By including additional symmetries, all the
topological phases in terms of Majorana fermions have been classified within
the matrix product representation\cite{Bultinck2017,Kapustin2016}. If the
MPS for topological phases with parafermions are constructed, we have to
generalize the concept of fermionic parity and establish the associated
basis states.

In this paper, we introduce an intuitive \textquotedblright
particle-like\textquotedblright\ representation of parafermions in real space%
\cite{CobaneraOrtiz}, the generalization of Majorana fermions in the Fock
space. By a local transformation, these Fock parafermions are connected to
the (Weyl) parafermions introduced from the $\mathbb{Z}_{p}$ spin degrees of
freedom ($p$ is an integer number). The indistinguishable Fock parafermions
satisfy the correlated $p$-exclusion and $2\pi /p$-exchange statistics\cite%
{CobaneraOrtiz}. In the Fock parafermion representation, a natural
formulation of the generalized MPS for topological phases with parafermions
can be established. For $p=3$, only two topologically distinct classes of
irreducible parafermionic MPS can be constructed, characterized by the
presence of one or two parafermionic zero modes at each end of open chains.
The derived parent Hamiltonians are found as the fixed point lattice models
of the single $\mathbb{Z}_{3}$ parafermion chain\cite{Fendley2012} and
two-coupled parafermion chains with $\mathbb{Z}_{3}\times \mathbb{Z}_{3}$
symmetry\cite{Pollmann2013}. For the topological parafermionic MPS states
protected by the $\mathbb{Z}_{3}\times \mathbb{Z}_{3}$ symmetry, there also
exists two nontrivial distinct classes, resulting from two different ways to
stack the MPS wave functions for two separate $\mathbb{Z}_{3}$ parafermion
chains. So our general framework can be easily generalized to construct the $%
\mathbb{Z}_{p}$ symmetric MPS for one-dimensional topological phases. When
additional symmetries are included into the $\mathbb{Z}_{p}$ parafermionic
chains, we can classify all the possible topological phases. Moreover, our
present formulation can also be used to construct the tensor product states
for topological phases in more than one spatial dimension\cite%
{Gu-Verstraete-Wen,Eisert,VerstraeteMPO}.

In Sec.II, we discuss the Fock space of parafermions and their relations to
the Weyl parafermions. Then in Sec. III, we outline the general framework to
construct the MPS wave functions in the Fock representation of parafermions.
In Sec. IV, we explicitly derive the MPS wave functions for the $\mathbb{Z}%
_{3}$ parafermion chain and two-coupled $\mathbb{Z}_{3}\times \mathbb{Z}_{3}$
symmetric parafermion chains with various boundary conditions. Conclusion
and outlook are given in Sec. V.

\section{Fock parafermions}

It is known that, from the $\mathbb{Z}_{p}$ spin degrees of freedom of the
clock models, the parafermions are usually defined by a generalized
Jordan-Wigner transformation as\cite{FradkinKadonoff,AlcarazKoberle}
\begin{equation}
\chi _{2l-1}=\left( \prod_{k<l}\tau _{k}\right) \sigma _{l},\quad \chi
_{2l}=-e^{i\pi /p}\left( \prod_{k\leq l}\tau _{k}\right) \sigma _{l},
\end{equation}%
where the $\mathbb{Z}_{p}$ spin matrices are given by%
\begin{eqnarray}
\tau _{l} &=&\left(
\begin{array}{ccccc}
0 & 0 & \cdots & 0 & 1 \\
1 & 0 & \cdots & 0 & 0 \\
0 & 1 & \cdots & 0 & 0 \\
\vdots & \vdots & \ddots & 0 & 0 \\
0 & 0 & \cdots & 1 & 0%
\end{array}%
\right) ,  \notag \\
\sigma _{l} &=&\text{diag}\left( 1,\omega ,...,\omega ^{p-2},\omega
^{p-1}\right) ,
\end{eqnarray}%
and $\omega =e^{i2\pi /p}$ ($\bar{\omega}=e^{-i2\pi /p}$). From the
relations satisfied by the $\mathbb{Z}_{p}$ spin operators
\begin{equation}
\sigma _{l}^{p}=\tau _{l}^{p}=1,\quad \sigma _{l}\tau _{m}=\omega ^{\delta
_{l,m}}\tau _{m}\sigma _{l},
\end{equation}%
the algebra of the parafermions are determined as
\begin{equation}
\chi _{l}^{p}=1,\quad \chi _{l}\chi _{m}=\omega \chi _{m}\chi _{l},\quad
\text{for }l<m.
\end{equation}%
Since the algebra realized by the parafermions is the generalized Clifford
algebra first noticed by Weyl\cite{Weyl1950}, such parafermions are referred
to as Weyl parafermions. When $p=2$, they are just Majorana fermions.

It is known that the combinations of two Majorana fermions form one Dirac
fermion, and the states of Fock space are endowed with the parity of fermion
number, so the creation and annihilation operators can be defined
systematically. What is the second quantized description of the Weyl
parafermions? It is given by the Fock parafermions\cite{CobaneraOrtiz}. With
the basis of orthogonal single-particle orbitals: $\phi _{1}$,$\phi _{2}$,$%
...,\phi _{L}$, the many-body states of $\mathbb{Z}_{p}$ Fock parafermions
can be assumed as $|i_{1}i_{2}\cdots i_{L}\rangle $, where $i_{1}$, $i_{2}$,
$\cdots $, $i_{L}$ are the respective occupation numbers of the single
particle orbitals and $i_{l}\in \mathbb{Z}_{p}\equiv \left\{ 0,1,2,\cdots
,p-1\right\} $. The general structure of the Fock space can be defined by
\begin{equation}
\mathbb{V}_{F}=\bigoplus_{M=0}^{L(p-1)}\text{Span}\left\{ |i_{1}i_{2}\cdots
i_{L}\rangle ,|\sum_{l=1}^{L}i_{l}=M\right\} .
\end{equation}%
In the following we use the abbreviated notation $|i_{l}\rangle =|0\cdots
i_{l}\cdots 0\rangle $ for single-particle states. By considering the
non-trivial statistics of Weyl parafermions, the graded tensor product $%
\otimes _{g}$ should be introduced when constructing the many-body states
from the single-particle states,
\begin{eqnarray}
\langle i_{1}i_{2}\cdots i_{L}| &=&\langle i_{L}|\otimes _{g}\cdots \otimes
_{g}\langle i_{2}|\otimes _{g}\langle i_{1}|,  \notag \\
|i_{1}i_{2}\cdots i_{L}\rangle &=&|i_{1}\rangle \otimes _{g}|i_{2}\rangle
\otimes _{g}\cdots \otimes _{g}|i_{L}\rangle ,
\end{eqnarray}%
which is the exact mathematical description of the graded structure of
Hilbert space in a non-commuting system. The crucial ingredient of the
graded tensor product is the following isomorphism $\mathcal{F}$:
\begin{eqnarray}
\mathcal{F}(|i_{l}\rangle \otimes _{g}|j_{m}\rangle ) &\equiv &\omega
^{ij}|j_{m}\rangle \otimes _{g}|i_{l}\rangle ,  \notag \\
\mathcal{F}(\langle i_{l}|\otimes _{g}|j_{m}\rangle ) &\equiv &\bar{\omega}%
^{ij}|j_{m}\rangle \otimes _{g}\langle i_{l}|,
\end{eqnarray}%
for $l<m$. These multiplication rules capture the correlated $2\pi /p$%
-exchange statistics of Fock parafermions, which is the crucial point for
the construction of the MPS wave functions.

Since the orthogonality
\begin{equation}
\langle i_{1}i_{2}\cdots i_{L}|j_{1}j_{2}\cdots j_{L}\rangle
=\prod_{l=1}^{L}\delta _{i_{l},j_{l}},
\end{equation}%
is required, the contraction $\mathcal{C}$ has to be defined via a mapping $%
\mathbb{V}_{F}^{\ast }\otimes _{g}\mathbb{V}_{F}\rightarrow \mathbb{C}$,
which acts as
\begin{equation}
\mathcal{C}\left( \langle i_{l}|\otimes _{g}|j_{l}\rangle \right) =\langle
i_{l}|j_{l}\rangle =\delta _{i_{l},j_{l}}.
\end{equation}%
With the $2\pi /p$-exchange statistics for different orbitals and the
orthogonality relation as well as the property of vacuum state, one can
derive the $p$-exclusive principle for the same orbital
\begin{equation}
(|i_{l}=1\rangle )^{\otimes _{g}p}\equiv |i_{l}=p\rangle =0.
\end{equation}%
So the dimension of the Fock space of parafermions is determined as $p^{L}$.

Moreover, the creation operator $C_{l}^{\dagger }$ of Fock parafermions can
be introduced by
\begin{eqnarray}
C_{l}^{\dagger }|i_{1}\cdots i_{L}\rangle &=&|i_{l}=1\rangle \otimes
_{g}|i_{1}\cdots i_{L}\rangle  \notag \\
&=&\bar{\omega}^{\sum_{k<l}i_{k}}|i_{1}\cdots i_{l}+1\cdots ,i_{L}\rangle ,
\end{eqnarray}%
and the adjoint annihilation operator $C_{l}$ by
\begin{equation}
C_{l}|i_{1}\cdots i_{L}\rangle =\omega ^{\sum_{k<l}i_{k}}|i_{1}\cdots
i_{l}-1\cdots i_{L}\rangle .
\end{equation}%
The particle number operator is thus derived as%
\begin{equation}
{N}_{l}=\sum_{r=1}^{p-1}C_{l}^{\dagger r}C_{l}^{r}.
\end{equation}%
It can be easily proved that the creation and annihilation operators satisfy
the following relations
\begin{eqnarray}
C_{l}^{\dagger p} &=&0,\quad C_{l}^{\dagger }C_{m}^{\dagger }=\omega
C_{m}^{\dagger }C_{l}^{\dagger },  \notag \\
C_{l}^{p} &=&0,\quad C_{l}C_{m}=\omega C_{m}C_{l},  \notag \\
C_{l}^{\dagger }C_{m} &=&\bar{\omega}C_{m}C_{l}^{\dagger },\quad
C_{l}C_{m}^{\dagger }=\bar{\omega}C_{m}^{\dagger }C_{l},
\end{eqnarray}%
for different orbitals $l<m$, while for the same orbitals $l=m$, we have $%
p-1 $ relations
\begin{equation}
C_{l}^{\dagger r}C_{l}^{r}+C_{l}^{p-r}C_{l}^{\dagger \left( p-r\right)
}=1,\quad r=1,\cdots ,p-1.
\end{equation}%
For $p=2$, the above algebra for the creation and annihilation operators of
Fock parafermions will be reduced to the standard fermion algebra.

A natural question arises as what kind of expressions the Weyl parafermions
take in the Fock parafermion representation. It was Cobanera and Ortiz\cite%
{CobaneraOrtiz} who found the following remarkable relations%
\begin{equation}
\chi _{2l-1}=C_{l}+C_{l}^{\dagger \left( p-1\right) },\chi _{2l}=C_{l}\omega
^{{N}_{l}}+C_{l}^{\dagger \left( p-1\right) },
\end{equation}%
which generalize the standard relations between the Majorana fermions and
Dirac fermions. Formally we can still regard that the combination of two
Weyl parafermions forms one Fock parafermion. Their inverse transformations
can also be derived, giving rise to the local transformations between the
Fock parafermions and Weyl parafermions
\begin{eqnarray}
C_{l}^{\dagger } &=&\frac{p-1}{p}\chi _{2l-1}^{\dagger }-\frac{1}{p}%
\sum_{r=1}^{p-1}\overline{\omega }^{r(r+1)/2}\chi _{2l}^{r}\chi
_{2l-1}^{\dagger \left( r+1\right) },  \notag \\
C_{l} &=&\frac{p-1}{p}\chi _{2l-1}-\frac{1}{p}\sum_{r=1}^{p-1}\omega
^{r(r+1)/2}\chi _{2l-1}^{r+1}\chi _{2l}^{\dagger r}.
\end{eqnarray}%
Interestingly, these transformations are linear for $p=2$ but become
nonlinear for $p>2$.

Finally, a charge operator on a lattice site can be defined by $Q_{l}=\chi
_{2l-1}^{\dagger }\chi _{2l}=\omega ^{N_{l}}$, which yields the relation
between charge operator and the particle number operator of the Fock
parafermions. The global charge operator is thus given by
\begin{equation}
Q=\prod_{l}Q_{l}=\omega ^{\sum_{l}{N}_{l}},
\end{equation}%
and the global charge in the basis of the Fock space is determined by
\begin{equation}
Q|I\rangle =Q|i_{1}i_{2}\cdots i_{L}\rangle =\omega ^{i_{1}+i_{2}+\cdots
+i_{L}}|i_{1}i_{2}\cdots i_{L}\rangle .
\end{equation}%
Then the charge of the many-body basis state $|I\rangle $ denotes as $%
|I|=\left( \sum_{l=1}^{L}i_{l}\right) $ $\mathrm{mod}$ $p$, while the charge
of the basis state $\langle I|$ as $-|I|$. The tensor constructed by the
graded tensor product of Fock states with a definite charge has the total
charge given by the summation of the charges of these states ($\mathrm{mod}$
$p$). For example, the total charge of the tensor $|I\rangle \otimes
_{g}\langle J|$ is given by $\left( |I|-|J|\right) $ $\mathrm{mod}$ $p$.
Since the $\mathbb{Z}_{p}$ parafermionic system with $p$ as a prime number
always preserves the $\mathbb{Z}_{p}$ symmetry\cite{Quella}, which is
generated by the total charge operator $Q$, we expect that every
parafermionic many-body state formed by a linear superposition of the same
charge states should have a definite charge.

\section{MPS in the Fock representation}

\begin{figure}[tbp]
\centering 
\includegraphics[width=3.2in]{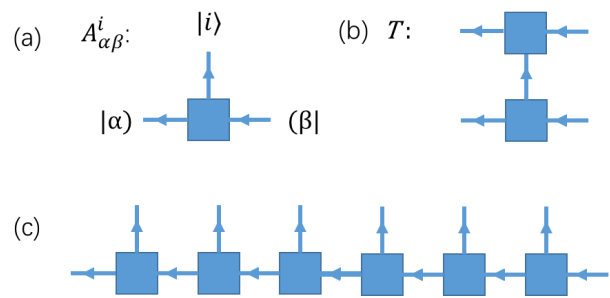}\newline
\caption{(a) The tensor used to construct MPS, the arrow pointing
inward/outward from the box represents the bra/ket state, respectively. (b)
Graphical representation of transfer matrix. (c) MPS constructed by
contracting the virtual states between the adjacent tensors without fixing
the boundaries.}
\label{1}
\end{figure}

To construct the MPS for physical degrees of freedom in dimension $d$, we
have to introduce two auxiliary virtual degrees of freedom in dimension $D$.
Two virtual degrees of freedom on the neighboring sites form a maximally
entangled state, and two auxiliary virtual degrees of freedom on the same
site are projected onto the physical degrees of freedom. Such a picture
captures the most important entanglement property of one-dimensional systems
satisfying the area law theorem. With the Fock parafermions, we can write
down the local tensor
\begin{equation}
\mathbf{A}[l]=\sum_{\alpha \beta i}A[l]_{\alpha \beta }^{i}|\alpha
_{l})\otimes _{g}|i_{l}\rangle \otimes _{g}(\beta _{l+1}|,
\label{LocalTensor}
\end{equation}%
where $l$ denotes the site index and $|\alpha _{l})$ and $(\beta _{l+1}|$
stand for the virtual states with the charges $|\alpha |,-|\beta |\in
\mathbb{Z}_{p}$ respectively, while $|i_{l}\rangle $ for the physical state
with the charge $|i|\in \mathbb{Z}_{p}$ in the Fock space. We can
graphically represent $\mathbf{A}[l]$ as shown in Fig.\ref{1}(a). It is the
tensor $\mathbf{A}[l]\in \mathbb{V}_{l}\otimes _{g}\mathbb{H}_{l}\otimes _{g}%
\mathbb{V}_{l+1}^{\ast }$ that maps from the virtual Hilbert space to the
physical Hilbert space. Sometimes we neglect the symbol $\otimes _{g}$ and
simply write $\mathbf{A}[l]=\sum_{\alpha ,\beta ,i}A[l]_{\alpha \beta
}^{i}|\alpha _{l})|i_{l}\rangle (\beta _{l+1}|$. One may think that a single
Fock parafermion is composed of two Weyl parafermions, so the virtual
degrees of freedom should be expressed by Weyl parafermions. Although we
still use the Fock representation to express the virtual degrees of freedom,
these degrees of freedom of the virtual space are nevertheless
fractionalized and describe fractionalized parafermionic zero modes at the
ends of open chains. Because the virtual states $|\alpha )$ and $(\beta |$
are not independent, there is a constrain of a fixed charge value $%
|i|+|\alpha |-|\beta |$ for a local tensor. In other words, since we are
unable to use the fractionalized virtual degrees of freedom $\sqrt{p}$ to
construct the MPS, we have to employ the unfractionalized degrees of freedom
with a constrain of charge to restrict the degrees of freedom. For example,
when the fermionic MPS for the single Majorana fermion chain with open
boundary conditions are constructed, there are four different choices to fix
the virtual degrees of freedom on two edges, but only two-fold ground state
degeneracy is produced\cite{Bultinck2017}.

To manipulate the tensor network locally, we will always impose the
constraint that all the local tensors $\mathbf{A}[l]$\ must have a
well-defined charge, so that the different orders of the tensors in the
graded tensor product will at most lead to a global phase of the many-body
state. To ensure that the tensor $\mathbf{A}[l]$ has a definite charge, we
must use the matrices $\mathbf{A}[l]^{i}$ with a well-defined charge. We
mainly focus on the translational invariant systems with the same $\mathbf{A}%
[l]$ on different sites and the site index will be omitted in the following.
Then we impose that the charge of the local tensor $\mathbf{A}$ is zero so
that the order of the individual tensors $\mathbf{A}$ does not affect the
definition of tensor network and its charge is independent on the lattice
size. When we contract the virtual bonds by the mapping $\mathcal{C}$, we
have the general parafermionic MPS as
\begin{eqnarray}
|\psi \rangle &=&\mathcal{C}(\mathbf{A}[1]\otimes _{g}\mathbf{A}[2]\otimes
_{g}\cdots \otimes _{g}\mathbf{A}[L]) \\
&=&\sum_{i_{1}..i_{N}}\sum_{\alpha \beta }\left( A^{i_{1}}\cdots
A^{i_{L}}\right) _{\alpha \beta }|\alpha _{1})|i_{1}\cdots i_{L}\rangle
(\beta _{L+1}|,  \notag
\end{eqnarray}%
which has been shown in Fig. \ref{1}(c). Because the charges of $|\alpha
_{l})$ and $(\alpha _{l}|$ add up to zero, the contraction does not affect
the charge of the state $|\psi \rangle $. Finally, we can choose different
boundary conditions to fix the virtual states at the boundaries. Because the
commutation relation of parafermions requires a definite order, the
translation invariance on a closed chain is not well-defined.

For any MPS wave function, one can construct a local parent Hamiltonian
whose ground state is uniquely given by the MPS. The general way to
construct the parent Hamiltonian is as follows. First we should block the
consecutive $m$ sites so that the rank of the map $B_{\alpha \beta
}^{i}=(A^{i_{1}}A^{i_{2}}\cdots A^{i_{m}})_{\alpha \beta }$ with row indices
$i$ and the column indices $\alpha \beta $ is smaller than $d^{m}$. Then a
local Hamiltonian is obtained by projecting onto the space orthogonal to the
image of this map. In physics, the image of this map is the ground state
subspace. Since the operator $P=BB^{+}$ is the orthogonal projector onto the
image of $B$, where $B^{+}$ is the Moore-Penrose pseudo-inverse of the
matrix $B$, we can construct a gapped frustration free parent Hamiltonian by
adding all local projectors
\begin{equation}
H=\sum_{l}(1-P_{l}),
\end{equation}%
where $1-P_{l}$ is the orthogonal projector onto the Hilbert space
orthogonal to the image of $B$, the kernel of $B^{+}$. Then, if the local
Hamiltonian matrix is simple, i.e., it only involves a few number of lattice
sites, we can try to expand it with possible charge zero operators in terms
of the parafermion modes defined on those lattice sites. Neglecting the
unnecessary terms, the simplified parent Hamiltonian corresponding to the
constructed MPS wave function is thus obtained.

\section{$\mathbb{Z}_{3}$ parafermionic MPS}

Following the general procedures of constructing MPS with $\mathbb{Z}_{3}$
Fock parafermions, the charges of elements of the local matrices $\mathbf{A}%
^{i}=\sum_{\alpha \beta }A_{\alpha \beta }^{i}|\alpha _{l})\otimes
_{g}(\beta _{l+1}|$ (the components of the local tensor) are determined by $%
(\beta -\alpha )$ $\mathrm{mod}$ $3$, shown in Tab.\ref{table1}. To ensure
that the local matrix $\mathbf{A}^{i}$ has a definite charge, we require
that the matrix form of $A_{\alpha \beta }^{i}$ must have the following nine
blocks structure:
\begin{eqnarray}
A^{i} &=&\left(
\begin{array}{ccc}
a_{0} & 0 & 0 \\
0 & b_{0} & 0 \\
0 & 0 & c_{0}%
\end{array}%
\right) ,\quad \text{if\ }|i|=0\text{,}  \notag \\
\text{ }A^{i} &=&\left(
\begin{array}{ccc}
0 & a_{1} & 0 \\
0 & 0 & b_{1} \\
c_{1} & 0 & 0%
\end{array}%
\right) ,\quad \text{if\ }|i|=1\text{,}  \notag \\
\text{\ }A^{i} &=&\left(
\begin{array}{ccc}
0 & 0 & a_{2} \\
b_{2} & 0 & 0 \\
0 & c_{2} & 0%
\end{array}%
\right) ,\quad \text{if\ }|i|=2\text{,}
\end{eqnarray}%
which is determined by the graded structure of Fock space.
\begin{table}[tbp]
\begin{tabular}{|c|c|c|c|}
\hline
\diagbox{$|\alpha|$}{$|\mathbf{A}^{i}_{\alpha\beta}{[l]}|$}{$-|\beta|$} & 0
& 2 & 1 \\ \hline
0 & 0 & 2 & 1 \\ \hline
1 & 1 & 0 & 2 \\ \hline
2 & 2 & 1 & 0 \\ \hline
\end{tabular}%
\caption{The charges of different elements of $\mathbf{A}^{i}$, which reveal
the graded structure of $A^{i}$.}
\label{table1}
\end{table}

Actually these matrices form a family of $\mathbb{Z}_{3}$ symmetric states,
and all $A^{i}$ span a $\mathbb{Z}_{3}$ graded algebra. If we multiply some
of these matrices $A^{i},A^{j},\cdots ,A^{k}$ together to get a blocked
matrix $B^{m}=A^{i}A^{j}\cdots A^{k}$ with $|m|=\left( |i|+|j|+\cdots
+|k|\right) $ $\mathrm{mod}$ $3$, the blocked matrix $B^{m}$ still satisfies
the above conditions. The $\mathbb{Z}_{3}$ graded algebra is the consequence
that the system must be $\mathbb{Z}_{3}$ symmetric, so the $\mathbb{Z}_{3}$
charge conserves and the states in the Hilbert space must have a
well-defined charge. Hence the MPS with different charges cannot be
connected smoothly. Since the $\mathbb{Z}_{3}$ parafermionic system is
intrinsically $\mathbb{Z}_{3}$ symmetric, the Hilbert space is naturally a $%
\mathbb{Z}_{3}$ graded vector space, which is a generalization of super
vector space for the $\mathbb{Z}_{2}$ case\cite{Bultinck2017,Kapustin2016}.

\subsection{Wave functions and parent Hamiltonian}

A prototype parafermionic MPS can be defined by the simplest matrices
\begin{equation*}
A^{0}=\left(
\begin{array}{ccc}
1 & 0 & 0 \\
0 & 1 & 0 \\
0 & 0 & 1%
\end{array}%
\right) ,A^{1}=\left(
\begin{array}{ccc}
0 & 1 & 0 \\
0 & 0 & 1 \\
1 & 0 & 0%
\end{array}%
\right) ,A^{2}=\left(
\begin{array}{ccc}
0 & 0 & 1 \\
1 & 0 & 0 \\
0 & 1 & 0%
\end{array}%
\right) .
\end{equation*}%
The MPS with open boundary conditions constructed by these matrices can be
expressed as
\begin{equation}
|\psi _{n}\rangle =\sum_{\{i_{l}\}}(A^{i_{1}}A^{i_{2}}\cdots
A^{i_{L}})_{\alpha \beta }|i_{1}i_{2}\cdots i_{L}\rangle ,
\end{equation}%
where $n=\left( \alpha -\beta \right) $ $\mathrm{mod}$ $3$. Although we have
nine different choices of ($\alpha ,\beta $) corresponding to nine different
boundary conditions, there are only three topologically distinct ground
states carrying three $\mathbb{Z}_{3}$ charges $n=0,1,2$, because the $%
\mathbb{Z}_{3}$ graded structure of the matrices of $A^{i}$ shown in the
Tab. \ref{table1}. Specifically, they are equal wight superpositions of all
basis states with the same charge
\begin{equation}
|\psi _{n}\rangle =\sum_{\left( \Sigma _{l}i_{l}\right) \text{ }\mathrm{mod}%
\text{ }3\text{ }=n}|i_{1}i_{2}\cdots i_{L}\rangle ,\text{ }n=0,1,2.
\end{equation}%
Here we would like to emphasize that the summation has to satisfy the global
constrain. Similar MPS state is also considered in the recent paper\cite%
{Mazza}.

With the parafermionic MPS wave functions, the transfer matrix $%
T=\sum_{i}A^{i}\otimes \bar{A}^{i}$ which is shown graphically in Fig. \ref%
{1}(b) can be defined, and we can calculate its eigenvalue spectrum. We
surprisingly found that there is a unique eigenvalue with three-fold
degeneracy, indicating the distinct property of the parafermionic MPS. In
the bosonic MPS, however, the degeneracy of the largest eigenvalue of the
transfer matrix stems from the spontaneous symmetry breaking in the ground
state. But the $\mathbb{Z}_{3}$ symmetry can not be spontaneously broken in
the parafermion chain. This result becomes the most striking difference
between the bosonic and parafermionic MPS. In this sense, the above matrices
$A^{0},A^{1}$ and $A^{2}$ just form a \textquotedblleft
non-trivial\textquotedblright\ type of the $\mathbb{Z}_{3}$ graded algebra,
which has a non-trivial center formed by these three matrices.

Using the method of deriving the parent Hamiltonian, we can obtain the model
Hamiltonian, corresponding to the fixed-point Hamiltonian for the
non-trivial phase of a single parafermion chain\cite{Fendley2012}
\begin{eqnarray}
H &=&-\sum_{l=1}^{L-1}\left[ \bar{\omega}^{{N}_{l}-1}\left( C_{l}^{\dagger
}C_{l+1}+C_{l}^{\dagger }C_{l+1}^{\dagger 2}\right) \right.  \notag \\
&&\text{ \ \ \ }\left. +\omega \left(
C_{l}^{2}C_{l+1}+C_{l}^{2}C_{l+1}^{\dagger 2}\right) +\text{h.c.}\right]
\notag \\
&=&-\sum_{l=1}^{L-1}\left( \omega \chi _{2l}^{\dagger }\chi _{2l+1}+\bar{%
\omega}\chi _{2l+1}^{\dagger }\chi _{2l}\right) ,
\end{eqnarray}%
where we have used the relations between the Fock parafermions and Weyl
parafermions
\begin{eqnarray}
C_{l} &=&\frac{2}{3}\chi _{2l-1}-\frac{1}{3}\left( \chi _{2l}+\omega \chi
_{2l-1}^{\dagger }\chi _{2l}^{\dagger }\right) ,  \notag \\
C_{l}^{\dagger } &=&\frac{2}{3}\chi _{2l-1}^{\dagger }-\frac{1}{3}\left(
\chi _{2l}^{\dagger }+\bar{\omega}\chi _{2l}\chi _{2l-1}\right) .
\end{eqnarray}%
In the Fock representation, we notice that the parent Hamiltonian includes
the nearest neighbor single-particle and two-particle hopping terms as well
as the three-parafermion \textquotedblleft pairing\textquotedblright\ terms
on the nearest neighbor sites. However, in terms of Weyl parafermions, the
parent Hamiltonian just takes a simple form, the nearest neighbor hopping
terms, and we can easily find that the Weyl parafermions $\chi _{1}^{\dagger
}$ and $\chi _{2L}$ characterize two edge parafermion zero modes on each end
of the open chain. These edge parafermions are fractionalized from the
physical degrees of freedom on the lattice sites. Since we can not
distinguish three ground states $|\psi _{0}\rangle ,|\psi _{1}\rangle ,|\psi
_{2}\rangle $ locally in the bulk, they are topologically non-trivial
degenerate states. In addiction, via the generalized inverse Jordan-Wigner
transformation, one can transform the above parent Hamiltonian $H$ into the
ferromagnetic $\mathbb{Z}_{3}$ clock model, whose ground state has a
long-range order with three-fold degeneracy due to the spontaneous $\mathbb{Z%
}_{3}$ symmetry breaking.

Since the coupling parameters can be rotated by an angle $\pm 2\pi /3$, we
further noticed that there are two other equivalent parent Hamiltonians\cite%
{H H Tu}
\begin{eqnarray}
H^{\prime } &=&-\sum_{l=1}^{L-1}\left( \chi _{2l}^{\dagger }\chi
_{2l+1}+\chi _{2l+1}^{\dagger }\chi _{2l}\right) ,  \notag \\
H^{\prime \prime } &=&-\sum_{l=1}^{L-1}\left( \bar{\omega}\chi
_{2l}^{\dagger }\chi _{2l+1}+\omega \chi _{2l+1}^{\dagger }\chi _{2l}\right)
,
\end{eqnarray}%
whose charges of their corresponding MPS depend on the lattice size. This
can be seen from the total charge operator
\begin{eqnarray*}
Q &=&\chi _{1}^{\dagger }\chi _{2}\chi _{3}^{\dagger }\chi _{4}\cdots \chi
_{2L-1}^{\dagger }\chi _{2L} \\
&=&\chi _{1}^{\dagger }\left( \bar{\omega}\chi _{3}^{\dagger }\chi
_{2}\right) \left( \bar{\omega}\chi _{5}^{\dagger }\chi _{4}\right) \cdots
\left( \bar{\omega}\chi _{2L-1}^{\dagger }\chi _{2L-2}\right) \chi _{2L},
\end{eqnarray*}%
where the parafermionic operators on the adjacent sites define the
\textquotedblleft bond charge\textquotedblright\ $\bar{\omega}\chi
_{2l+1}^{\dagger }\chi _{2l}$. Three degenerate ground states with zero bond
charge can minimize the energy of the parent Hamiltonian $H$, while the
ground states with \textquotedblleft bond charge\textquotedblright\ $2$ and $%
1$ can make the energies of $H^{\prime }$ and $H^{\prime \prime }$ minimize,
respectively. Those MPS wave functions must be approximated by the local
tensors $\mathbf{A}$ with charge $2$ and $1$, respectively.

\begin{figure}[tbp]
\centering 
\includegraphics[width=3.2in]{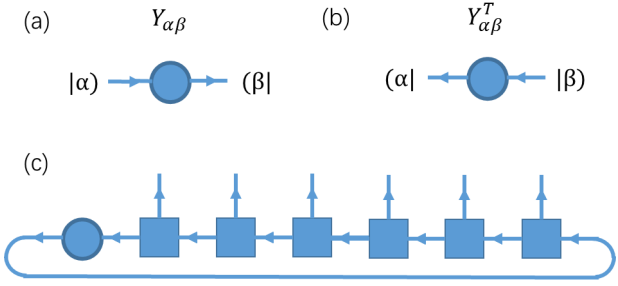}\newline
\caption{(a) The closure tensor $\mathbf{Y}$. (b) The transpose closure
tensor $\mathbf{Y}^{T}$. (c) Parafermionic MPS with closed boundary
condition.}
\label{3}
\end{figure}

Next we consider the corresponding parafermionic MPS in the closed boundary
conditions. A closure tensor $\mathbf{Y}_{a}$ with the charge $a=0,1,2$ can
be introduced in the wave function as follows
\begin{eqnarray}
|\psi _{a}\rangle &=&\mathcal{C}(\mathbf{Y}_{a}\mathbf{\otimes }_{g}\mathbf{A%
}[1]\otimes _{g}\mathbf{A}[2]\otimes _{g}\cdots \otimes _{g}\mathbf{A}[L])
\notag \\
&=&\sum_{\{i_{l}\}}\text{tr}(Y_{a}^{T}A^{i_{1}}A^{i_{2}}\cdots
A^{i_{L}})|i_{1}i_{2}\cdots i_{L}\rangle ,
\end{eqnarray}%
where $\mathbf{Y}_{a}\mathbf{=}\sum_{\gamma \delta }Y_{a,\gamma \delta
}(\gamma _{1}|\otimes _{g}|\delta _{L})$. The closure tensor $Y$ and the
closed wave function are shown in Fig. \ref{3}. Since all local tensors $%
\mathbf{A}$ have charge $0$, the moving of $|\delta _{L})$ to the right end
does not bring any phase. Different choices of $\mathbf{Y}_{a}$ just result
in the different charges of the closed wave functions in the view of the
\textquotedblleft boundary charge\textquotedblright\ as shown
\begin{equation}
Q=\chi _{1}^{\dagger }\chi _{2}\chi _{3}^{\dagger }\chi _{4}\cdots \chi
_{2L-1}^{\dagger }\chi _{2L}=\bar{\omega}\left( \chi _{2L}\chi _{1}^{\dagger
}\right) .
\end{equation}%
Therefore, we can write the parent Hamiltonians corresponding to the MPS
with the closed boundary conditions
\begin{equation}
H^{(a)}=H-\left( \omega ^{a}\chi _{2L}^{\dagger }\chi _{1}+\bar{\omega}%
^{a}\chi _{1}^{\dagger }\chi _{2L}\right) ,\text{ }a=0,1,2.
\end{equation}%
Actually the translational invariance of a closed parafermion chain is very
tricky. It seems that $H^{(1)}$ is translational invariant, but from the
"bond charge" point of view, it corresponds to the MPS with a charge $2$. By
applying the translation operator to the MPS wave functions, we find that
none of them satisfies the periodic boundary condition\cite{Bultinck2017}.
We attribute this to the ordering requirement in the commutation relation of
parafermions. From topological bulk response, however, we know that three
different boundary conditions can select three different ground states under
the open boundary conditions. So we immediately identify
\begin{equation*}
Y_{0}=\left(
\begin{array}{ccc}
1 & 0 & 0 \\
0 & 1 & 0 \\
0 & 0 & 1%
\end{array}%
\right) ,Y_{1}=\left(
\begin{array}{ccc}
0 & 1 & 0 \\
0 & 0 & 1 \\
1 & 0 & 0%
\end{array}%
\right) ,Y_{2}=\left(
\begin{array}{ccc}
0 & 0 & 1 \\
1 & 0 & 0 \\
0 & 1 & 0%
\end{array}%
\right) .
\end{equation*}%

\subsection{Stacking two $\mathbb{Z}_{3}$ parafermionic MPS}

According to the classification of the $\mathbb{Z}_{p}$ symmetric
parafermion chains with a prime integer $p$, there only exist \textit{two}
topologically distinct gapped phases\cite{Pollmann2013,Quella}, including a
topological nontrivial phase and trivial gapped phase. We have just
constructed one topologically nontrivial $\mathbb{Z}_{3}$ symmetric MPS, and
another\ $\mathbb{Z}_{3}$ "trivial" type algebra of the parafermionic MPS
wave function has not been considered. Similar to the $\mathbb{Z}_{2}$ case%
\cite{Bultinck2017,Kapustin2016}, stacking two separate parafermionic MPS in
one unit cell will automatically gives rise to the other type parafermionic
MPS. We start from the graded tensor product of two different local tensors
and record the induced phases. For the local tensors $\mathbf{A}$ with
charge $0$, we have
\begin{eqnarray}
&&\mathcal{F}(\mathbf{A}\otimes _{g}\mathbf{A}^{\prime })  \notag \\
&=&\mathcal{F}\left[ \sum_{i\alpha \beta }\left( A_{\alpha \beta
}^{i}|\alpha )|i\rangle (\beta |\right) \otimes _{g}\sum_{j\gamma \delta
}\left( A_{\gamma \delta }^{\prime j}|\gamma )|j\rangle (\delta |\right) %
\right] \text{\ \ }  \notag \\
&=&\sum_{ij\alpha \beta \gamma \delta }A_{\alpha \beta }^{i}A_{\gamma \delta
}^{\prime j}\omega ^{|\gamma ||i||}|\alpha )|\gamma )|i\rangle |j\rangle
(\delta |(\beta |.
\end{eqnarray}%
Then new local matrices building up the stacking MPS are defined by
\begin{equation}
B_{(\alpha \gamma )(\beta \delta )}^{ij}=A_{\alpha \beta }^{i}A_{\gamma
\delta }^{\prime j}\omega ^{|\gamma ||i|}.
\end{equation}%
With the matrices for the single parafermion chain, we have nine local
matrices with the smaller indices arranged before the larger indices. The
resulting MPS wave function is
\begin{equation}
|\phi \rangle =\sum_{\{i_{l},j_{l}\}}\left(
B^{i_{1}j_{1}}B^{i_{2}j_{2}}\cdots B^{i_{L}j_{L}}\right) _{\alpha \beta
}|i_{1}j_{1}i_{2}j_{2}\cdots i_{L}j_{L}\rangle .
\end{equation}%
If we write down these composite MPS wave functions explicitly, there are $81
$ different boundary conditions ($\alpha ,\beta $). But only nine
topologically different states exist. We also find the entanglement spectrum
with a nine-fold degeneracy. Actually when the wave functions are explicitly
written down, one can find that the total charges of the even and odd
sublattices generated by the operators $Q_{1}=\omega ^{\sum_{i}N_{2i-1}}$
and $Q_{2}=\omega ^{\sum_{i}N_{2i}}$ are conserved separately. So the
stacking MPS wave functions display a larger $\mathbb{Z}_{3}\times $ $%
\mathbb{Z}_{3}$ symmetry.

Furthermore, according to the bosonic and fermionic MPS\cite{Bultinck2017},
it is expected that the parafermionic MPS is reducible when the local
matrices do not span a simple $\mathbb{Z}_{3}$ graded algebra. Via the
following\ gauge transformation
\begin{equation}
G=\frac{\sqrt{3}}{3}\left(
\begin{array}{ccccccccc}
\omega & 0 & 0 & \bar{\omega} & 0 & 0 & 1 & 0 & 0 \\
1 & 0 & 0 & \bar{\omega} & 0 & 0 & \omega & 0 & 0 \\
1 & 0 & 0 & 1 & 0 & 0 & 1 & 0 & 0 \\
0 & 0 & \omega & 0 & 0 & \bar{\omega} & 0 & 0 & 1 \\
0 & 0 & 1 & 0 & 0 & \bar{\omega} & 0 & 0 & \omega \\
0 & 0 & 1 & 0 & 0 & 1 & 0 & 0 & 1 \\
0 & \omega & 0 & 0 & \bar{\omega} & 0 & 0 & 1 & 0 \\
0 & 1 & 0 & 0 & \bar{\omega} & 0 & 0 & \omega & 0 \\
0 & 1 & 0 & 0 & 1 & 0 & 0 & 1 & 0%
\end{array}%
\right) ,
\end{equation}%
those local matrices can be transformed into nine block diagonal matrices,
and the phase difference in diagonal blocks just causes a global phase
change. Hence $B^{ij}$ can be reduced to nine $3\times 3$ matrices
\begin{eqnarray}
B^{00} &=&\left(
\begin{array}{ccc}
1 & 0 & 0 \\
0 & 1 & 0 \\
0 & 0 & 1%
\end{array}%
\right) ,B^{10}=\left(
\begin{array}{ccc}
0 & 0 & \bar{\omega} \\
\omega & 0 & 0 \\
0 & 1 & 0%
\end{array}%
\right) ,  \notag \\
B^{01} &=&\left(
\begin{array}{ccc}
0 & 0 & 1 \\
1 & 0 & 0 \\
0 & 1 & 0%
\end{array}%
\right) ,B^{11}=\left(
\begin{array}{ccc}
0 & \bar{\omega} & 0 \\
0 & 0 & \omega \\
1 & 0 & 0%
\end{array}%
\right) ,  \notag \\
B^{02} &=&\left(
\begin{array}{ccc}
0 & 1 & 0 \\
0 & 0 & 1 \\
1 & 0 & 0%
\end{array}%
\right) ,B^{12}=\left(
\begin{array}{ccc}
\bar{\omega} & 0 & 0 \\
0 & \omega & 0 \\
0 & 0 & 1%
\end{array}%
\right) ,  \notag \\
B^{20} &=&\left(
\begin{array}{ccc}
0 & \bar{\omega} & 0 \\
0 & 0 & 1 \\
\omega & 0 & 0%
\end{array}%
\right) ,B^{21}=\left(
\begin{array}{ccc}
\bar{\omega} & 0 & 0 \\
0 & 1 & 0 \\
0 & 0 & \omega%
\end{array}%
\right) ,  \notag \\
B^{22} &=&\left(
\begin{array}{ccc}
0 & 0 & \bar{\omega} \\
1 & 0 & 0 \\
0 & \omega & 0%
\end{array}%
\right) .
\end{eqnarray}%
These matrices form the \textquotedblleft trivial\textquotedblright\ type of
$\mathbb{Z}_{3}$ graded algebra, which is nothing but ungraded algebra,
because its center is an identity. For the closed boundary condition, there
is a unique ground state so that the two sublattice symmetries cannot be
broken. The final wave functions are also the superpositions of states with
the same sublattice charges.

By combining two closure matrices, we can find the closure matrix for the
stacking MPS. For example, two charge $1$ closures give rise to
\begin{eqnarray}
\mathbf{Y} &=&\mathbf{Y}_{1}\otimes _{g}\mathbf{Y}_{2}  \notag  \label{Y1} \\
&\mathbf{=}&\sum_{\alpha \beta }Y_{1,\alpha \beta }(\alpha _{1}|\otimes
_{g}|\beta _{1})\sum_{\gamma \delta }Y_{2,\gamma \delta }(\gamma
_{2}|\otimes _{g}|\delta _{2})  \notag \\
&=&\sum_{\alpha \beta \gamma \delta }Y_{(\alpha \gamma )(\beta \delta
)}(\gamma _{2}|(\alpha _{1}|\otimes _{g}|\beta _{1})|\delta _{2}),
\end{eqnarray}%
where $Y_{(\alpha \gamma )(\beta \delta )}=Y_{1,\alpha \beta }Y_{2,\gamma
\delta }\omega ^{|\gamma |}$. After the gauge transformation, $Y^{T}$ can be
reduced to
\begin{equation*}
Y^{T}=\left(
\begin{array}{ccc}
0 & 0 & 1 \\
\omega & 0 & 0 \\
0 & \bar{\omega} & 0%
\end{array}%
\right) ,
\end{equation*}%
and the corresponding MPS wave function is thus given by
\begin{equation*}
|\phi \rangle =\sum_{\{i_{l},j_{l}\}}\text{tr}%
(Y^{T}B^{i_{1}j_{1}}B^{i_{2}j_{2}}\cdots
B^{i_{L}j_{L}})|i_{1}j_{1}i_{2}j_{2}\cdots i_{L}j_{L}\rangle .
\end{equation*}%
This is the total charge-$2$ MPS with total sublattice charge ones on each
sublattice. It should be pointed out that the ground state properties in the
reducing procedure do not change under the closed boundary conditions only.
In the open boundary conditions, however, the gauge transformation induces a
linear combination between the degenerate ground states, and then the
reduced MPS breaks the $\mathbb{Z}_{3}\times \mathbb{Z}_{3}$ symmetry.

Furthermore, we can also calculate the eigenvalues of the transfer matrix $%
T^{\prime }=\sum_{ij}B^{ij}\otimes \bar{B}^{ij}$, but its unique eigenvalue
is nondegenerate. The non-degenerate of the eigenvalue just suggests the
\textquotedblleft trivial\textquotedblright\ type of $\mathbb{Z}_{3}$ graded
algebra of $B^{ij}$, similar to the bosonic MPS for one-dimensional
topological phases. Thus, from the $\mathbb{Z}_{3}\times \mathbb{Z}_{3}$
symmetry point of view, the stacking MPS wave function should belong to the
family of SPT phases. This manifests from the following analysis. Requiring
\begin{eqnarray}
V_{1}B^{ij}V_{1}^{-1} &=&\sum_{i^{\prime }j^{\prime }}U_{ij,i^{\prime
}j^{\prime }}^{(1)}B^{i^{\prime }j^{\prime }},\text{ }  \notag \\
V_{2}B^{ij}V_{2}^{-1} &=&\sum_{i^{\prime }j^{\prime }}U_{ij,i^{\prime
}j^{\prime }}^{(2)}B^{i^{\prime }j^{\prime }}
\end{eqnarray}%
with $U^{(1)}=\sigma \otimes \mathbb{I}$ and $U^{(2)}=\mathbb{I}\otimes
\sigma $ as local unitary transformations on one unit cell and linear
representations of $\mathbb{Z}_{3}\times \mathbb{Z}_{3}$ generators, we can
find the projective representations of $Q_{1}$ and $Q_{2}$ as
\begin{equation}
V_{1}=\left(
\begin{array}{ccc}
0 & 0 & 1 \\
1 & 0 & 0 \\
0 & 1 & 0%
\end{array}%
\right) ,V_{2}=\left(
\begin{array}{ccc}
0 & \bar{\omega} & 0 \\
0 & 0 & 1 \\
\omega & 0 & 0%
\end{array}%
\right) ,
\end{equation}%
from which we find the important relation%
\begin{equation}
V_{1}V_{2}=\bar{\omega}V_{2}V_{1},
\end{equation}%
with $\bar{\omega}$ as the factor of the projective representation
characterizing this SPT phase\cite%
{Chen-Gu-Wen-2011,Chen-Gu-Wen-2011(2),Geraedts}.

\begin{figure}[tbp]
\centering 
\includegraphics[width=3.2in]{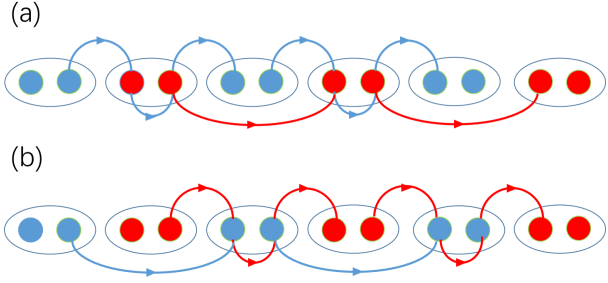}\newline
\caption{Graphical illustration of parent Hamiltonians $H_{2}$ (a) and $%
H_{2}^{\prime }$ (b). The ellipses denote different sites, blue and red dots
represent Weyl parafermions from original chain-1 and chain-2, respectively.
Blue and red lines denote the hopping terms originated from the hopping
terms of the chain-1 and chain-2, some of them keep quadratic but others
change into quartic form after combination.}
\label{twochainhamiltonian}
\end{figure}

Following the procedures to construct the parent Hamiltonian, we can derive
\begin{equation}
H_{2}=-\sum_{l=1}^{L-1}\left[ \omega \left( \chi _{4l}^{\dagger }\chi
_{4l+3}+\chi _{4l-2}\chi _{4l-1}\chi _{4l}^{\dagger }\chi _{4l+1}^{\dagger
}\right) +\text{h.c.}\right] ,
\end{equation}%
which commutes with $Q_{1}$ and $Q_{2}$. Note that the same parent
Hamiltonian is also obtained by using original matrices. Actually, $H_{2}$
can be viewed from the combination of two independent parafermion chains\cite%
{Pollmann2013}, as shown in the Fig. \ref{twochainhamiltonian}(a). Unlike
the $\mathbb{Z}_{2}$ Majorana fermion case, where the system after stacking
is two decoupled Kitaev chains and the edge modes are two single edge
Majorana operators, the parent Hamiltonian $H_{2}$ describes a two-coupled
parafermion chains satisfying the unusual commutation relation of Weyl
parafermions. By fractionalizing charge operators on two sublattices
separately\cite{Geraedts}, we can find that two parafermion zero modes exist
on each end of the two-coupled chains, given by the Weyl parafermions ($\chi
_{1}^{\dagger },\chi _{3},\chi _{4L-2}\chi _{4L-1}\chi _{4L}^{\dagger },\chi
_{4L}^{\dagger }$) for the chains with $2L$ sites and ($\chi _{1}^{\dagger
},\chi _{3},\chi _{4L}^{\dagger },\chi _{4L+2}$) for the chains with $2L+1$
sites. These edge Weyl parafermions commute with parent Hamiltonian and
carry $\mathbb{Z}_{3}$ charges. Two zero parafermion modes on the same edge
can actually form a physical degree of freedom three. In the case of $2L$
sites, due to the two edges are not symmetric, one of the edge mode is not a
single parafermion operator. The complicated edge modes and quartic
interactions in the parent Hamiltonian manifest the unusual commutation
relation of parafermions, which also endows the parafermion chain the
intrinsically strong-coupled property\cite{H H Tu}. The integer edge degrees
of freedom actually imply that there is no real fractionalization. Two edge
zero modes thus yield total nine-fold degenerate ground states for the open
chains. If we introduced the hoping term $\left( \chi _{1}^{\dagger }\chi
_{3}+\text{h.c.}\right) $ into the parent Hamiltonian, it would break the $%
\mathbb{Z}_{3}\times \mathbb{Z}_{3}$ symmetry and gap out the edge modes, so
we can thus prove that the edge modes are protected by $\mathbb{Z}_{3}\times
\mathbb{Z}_{3}$ symmetry. Actually there is a global unitary transformation%
\cite{Santos}
\begin{equation}
W=\exp \left[ \frac{i2\pi }{3}\sum_{l}\sum_{a=1}^{p-1}\frac{(\bar{\omega}%
\chi _{4l-2}^{\dag }\chi _{4l-1})^{a}-(\bar{\omega}\chi _{4l}^{\dag }\chi
_{4l+1})^{a}}{(\omega ^{a}-1)(\bar{\omega}^{a}-1)}\right] ,  \label{W}
\end{equation}%
which can transform the ground state of $H_{2}$ into a trivial gapped phase.
But such a unitary transformation explicitly breaks the $\mathbb{Z}%
_{3}\times \mathbb{Z}_{3}$ symmetry. Therefore, we conclude that the ground
state of $H_{2}$ is a SPT phase protected by the $\mathbb{Z}_{3}\times
\mathbb{Z}_{3}$ symmetry.

Moreover, in terms of $\mathbb{Z}_{3}$ spin operators via the inverse
generalized Jordan-Wigner transformation, this parent Hamiltonian can be
expressed as the $\mathbb{Z}_{3}\times \mathbb{Z}_{3}$ cluster model\cite%
{Santos,Geraedts}%
\begin{equation}
H_{2,\text{cl}}=-\sum_{l=1}^{L-1}\left( \sigma _{2l-1}\tau _{2l}\sigma
_{2l+1}^{\dagger }+\bar{\omega}\sigma _{2l}^{\dag }\tau _{2l+1}\sigma
_{2l+2}+\text{h.c}.\right) .
\end{equation}%
By redefinition of $\tau _{2l-1}$, we can wipe out $\bar{\omega}$ in front
of the second term. The ground state of $H_{2,\text{cl}}$ belongs to a SPT
phase protected by the $\mathbb{Z}_{3}\times \mathbb{Z}_{3}$ symmetry, which
can be proved from the view point of the bosonic matrix product
representation.

\subsection{Another way of stacking}

\begin{figure}[tbp]
\centering 
\includegraphics[width=3.2in]{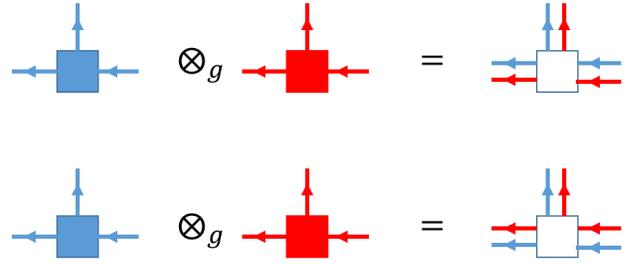}\newline
\caption{The blue/red graph represents the local tensor in the MPS of the
chain-1 or chain-2, and the differences are the recording order of the
virtual indices. In the second way of stacking, the index of the chain-1 is
larger than that of the chain-2 in the same unit cell.}
\label{4}
\end{figure}

Since the stacking MPS wave function displays the $\mathbb{Z}_{3}\times
\mathbb{Z}_{3}$ symmetry in bulk, there should exist another topologically
nontrivial gapped phases from the classification theory\cite%
{chen-gu-liu-wen,Santos}. In fact we can obtain another non-trivial SPT
phase by stacking of two single parafermion chains in a different recording
order:
\begin{eqnarray}
&&\mathcal{F}\left[ \sum_{i\alpha \beta }\left( A_{\alpha \beta }^{i}|\alpha
)|i\rangle (\beta |\right) \otimes _{g}\sum_{j\gamma \delta }\left(
A_{\gamma \delta }^{\prime j}|\gamma )|j\rangle (\delta |\right) \right]
\notag \\
&=&\sum_{ij\alpha \beta \gamma \delta }A_{\alpha \beta }^{i}A_{\gamma \delta
}^{\prime j}\omega ^{|j||\beta |}|\gamma )|\alpha )|i\rangle |j\rangle
(\beta |(\delta |,
\end{eqnarray}%
where we have assumed that the parafermion site-sequence number of the local
tensors $\mathbf{A}$ is larger than that of the local tensors $\mathbf{A}%
^{\prime }$. Two different recording ways are illustrated in the Fig. \ref{4}%
.

From the matrices given for the single parafermion chain, we have nine $%
9\times 9$ local matrices
\begin{equation}
D_{(\gamma \alpha )(\delta \beta )}^{ij}=A_{\alpha \beta }^{i}A_{\gamma
\delta }^{\prime j}\omega ^{|j||\beta |},
\end{equation}%
which can also be transformed into block diagonal matrix form by the gauge
transformation $G$. We then reduce these local matrices to%
\begin{eqnarray}
D^{00} &=&\left(
\begin{array}{ccc}
1 & 0 & 0 \\
0 & 1 & 0 \\
0 & 0 & 1%
\end{array}%
\right) ,D^{01}=\left(
\begin{array}{ccc}
0 & 0 & \bar{\omega} \\
\omega & 0 & 0 \\
0 & 1 & 0%
\end{array}%
\right) ,  \notag \\
D^{02} &=&\left(
\begin{array}{ccc}
0 & \bar{\omega} & 0 \\
0 & 0 & 1 \\
\omega & 0 & 0%
\end{array}%
\right) ,D^{10}=\left(
\begin{array}{ccc}
0 & 0 & 1 \\
1 & 0 & 0 \\
0 & 1 & 0%
\end{array}%
\right) ,  \notag \\
D^{11} &=&\left(
\begin{array}{ccc}
0 & 1 & 0 \\
0 & 0 & \bar{\omega} \\
\omega & 0 & 0%
\end{array}%
\right) ,D^{12}=\left(
\begin{array}{ccc}
\omega & 0 & 0 \\
0 & \bar{\omega} & 0 \\
0 & 0 & 1%
\end{array}%
\right) ,  \notag \\
D^{20} &=&\left(
\begin{array}{ccc}
0 & 1 & 0 \\
0 & 0 & 1 \\
1 & 0 & 0%
\end{array}%
\right) ,D^{21}=\left(
\begin{array}{ccc}
\omega & 0 & 0 \\
0 & 1 & 0 \\
0 & 0 & \bar{\omega}%
\end{array}%
\right) ,  \notag \\
D^{22} &=&\left(
\begin{array}{ccc}
0 & 0 & 1 \\
\omega & 0 & 0 \\
0 & \bar{\omega} & 0%
\end{array}%
\right) ,
\end{eqnarray}%
which also form the \textquotedblleft trivial\textquotedblright\ type $%
\mathbb{Z}_{3}$-graded algebra. As we pointed out, the $\mathbb{Z}_{3}\times
\mathbb{Z}_{3}$ symmetry is preserved for the MPS under the closed boundary
condition in the reducing procedure. For a closed boundary condition, the
corresponding MPS wave function can be expressed as%
\begin{equation*}
|\phi ^{\prime }\rangle =\sum_{\{i,j\}}\text{tr}%
(Y^{T}D^{i_{1}j_{1}}D^{i_{2}j_{2}}\cdots
D^{i_{L}j_{L}})|i_{1}j_{1}i_{2}j_{2}\cdots i_{L}j_{L}\rangle ,
\end{equation*}%
with the closure matrix $Y$ which can be derived as well. We notice that the
transfer matrix of $|\phi ^{\prime }\rangle $ is as same as that of $|\phi
\rangle $. One may think that they belong to the same phase, but in our
parafermionic case, they differ from each other in several aspects. For the
open boundary conditions, if we choose the simple boundary conditions such
as $(\alpha ,\beta )=(1,1)$, the $\mathbb{Z}_{3}\times \mathbb{Z}_{3}$
symmetry is broken and only $\mathbb{Z}_{3}$ symmetry is preserved, so both
of them belong to the $\mathbb{Z}_{3}$ symmetric trivial phase. Different
from the first stacking case, another projective representations of the $%
\mathbb{Z}_{3}\times \mathbb{Z}_{3}$ symmetry generators can be found
\begin{equation}
V_{1}^{\prime }=\left(
\begin{array}{ccc}
0 & \bar{\omega} & 0 \\
0 & 0 & 1 \\
\omega & 0 & 0%
\end{array}%
\right) ,V_{2}^{\prime }=\left(
\begin{array}{ccc}
0 & 0 & 1 \\
1 & 0 & 0 \\
0 & 1 & 0%
\end{array}%
\right) ,
\end{equation}%
with
\begin{equation}
V_{1}^{\prime }V_{2}^{\prime }=\omega V_{2}^{\prime }V_{1}^{\prime },
\end{equation}%
where the factor $\omega $ characterizes the MPS $|\phi ^{\prime }\rangle $
but the MPS $|\phi \rangle $ is characterized by $\bar{\omega}$. Both $|\phi
\rangle $ and $|\phi ^{\prime }\rangle $ represent two different
topologically nontrivial phases protected by the $\mathbb{Z}_{3}\times
\mathbb{Z}_{3}$ symmetry.

Furthermore, the corresponding parent Hamiltonian can also be found as
\begin{equation}
H_{2}^{\prime }=-\sum_{l=1}^{L-1}\left[ \omega \left( \chi _{4l-2}^{\dagger
}\chi _{4l+1}+\chi _{4l}\chi _{4l+1}\chi _{4l+2}^{\dagger }\chi
_{4l+3}^{\dagger }\right) +\text{h.c.}\right] ,
\end{equation}%
which is slightly different from the previous one $H_{2}$ and illustrated in
Fig.\ref{twochainhamiltonian} (b). This parent Hamiltonian also commutes
with $Q_{1}$ and $Q_{2}$. Similarly, by fractionalizing charge operators of
two sublattices, we can find the edge parafermion zero modes as ($\chi
_{1},\chi _{1}\chi _{2}^{\dagger }\chi _{3}^{\dagger },\chi _{4L-2}^{\dagger
},\chi _{4L}$) for the chain with $2L$ sites and ($\chi _{1},\chi _{1}\chi
_{2}^{\dagger }\chi _{3}^{\dagger },\chi _{4L}\chi _{4L+1}\chi
_{4L+2}^{\dagger },\chi _{4L+2}^{\dagger }$) for the chain with $2L+1$
sites, which commute with the parent Hamiltonian. In terms of $\mathbb{Z}%
_{3} $ spin operators, $H_{2}^{\prime }$ is further transformed into the
following $\mathbb{Z}_{3}\times \mathbb{Z}_{3}$ cluster model
\begin{equation}
H_{2,\text{cl}}^{\prime }=-\sum_{l=1}^{L-1}\left( \bar{\omega}\sigma
_{2l-1}^{\dagger }\tau _{2l}\sigma _{2l+1}+\sigma _{2l}\tau _{2l+1}\sigma
_{2l+2}^{\dag }+\text{h.c.}\right) ,
\end{equation}%
whose ground state represents another SPT phase protected by the $\mathbb{Z}%
_{3}\times \mathbb{Z}_{3}$ symmetry. By a global unitary transformation
without having $\mathbb{Z}_{3}\times \mathbb{Z}_{3}$ symmetry\cite{Santos}, $%
H_{2,\text{cl}}^{\prime }$ can be transformed into the previous one $H_{2,%
\text{cl}}$. The global unitary transformation can also be written in terms
of parafermions directly, similar to Eq.(\ref{W}). Again, we confirmed that
the ground states of $H_{2}$ and $H_{2}^{\prime }$ belong to different SPT
phases with the $\mathbb{Z}_{3}\times \mathbb{Z}_{3}$ symmetry.

\section{Conclusion and Outlook}

In the Fock representation of $\mathbb{Z}_{3}$ parafermions, we have
successfully constructed two topologically distinct classes of irreducible
parafermionic MPS, corresponding to the $\mathbb{Z}_{2}$ topological
classification of parafermion chain and the center of algebra spanned by the
local matrices. The nontrivial MPS represents the fixed point models of the
single $\mathbb{Z}_{3}$ parafermion chain, characterized by one parafermion
on each end of an open chain. The corresponding transfer matrix has a unique
eigenvalue with three-fold degeneracy, significantly different from that of
the bosonic MPS wave functions, where such a degeneracy implies the symmetry
breaking phase and the MPS is reducible. But in the parafermionic case, it
is irreducible and there is no symmetry breaking. The trivial type
parafermionic MPS wave functions obtaining by stacking of two single
parafermion chains have two edge parafermions on each end of an open chain
and usually display $\mathbb{Z}_{3}\times \mathbb{Z}_{3}$ symmetry. Hence
they are parafermionic SPT phases protected by $\mathbb{Z}_{3}\times \mathbb{%
Z}_{3}$ symmetry, much similar to the bosonic SPT phases for the two-coupled
$\mathbb{Z}_{3}\times \mathbb{Z}_{3}$ cluster models.

Our general framework can be easily generalized to construct the $\mathbb{Z}%
_{p}$ symmetric MPS for the one-dimensional parafermionic topological phases
with $p$ as a prime number, where the appearance of symmetry breaking in
non-prime $p$ cases gives rise to some complexity. When additional
symmetries are introduced into the $\mathbb{Z}_{3}$ parafermionic chains, we
can classify all the possible topological phases within the framework of the
MPS representation. Moreover, we can also generalize the tensor product
states for topological phases of parafermions in more than one spatial
dimension. Finally, we would like to emphasize that explorations of
topological phases with parafermions are not purely academic. Recently some
plausible experimental routes to trapping parafermionic excitations have
been proposed in presently available condensed matter systems, and these
platforms can provide topological qubits with both better protected against
environmental noise and richer fault-tolerant qubit rotations compared to
the Majorana-based systems\cite{AliceaFendley}.

\textit{Acknowledgment.- }The authors would like to thank Hong-Hao Tu for
his stimulating discussion and acknowledges the support of National Key
Research and Development Program of China (2016YFA0300300).

\end{document}